\documentclass[11pt]{article}

\usepackage{fullpage,times,amssymb}
\newtheorem{definition}{Definition} % Specify Definition
\newtheorem{fact}{Fact}    % Specify Fact
\newtheorem{theorem}{Theorem}    % Specify Theorem
\newtheorem{corollary}{Corollary}    % Specify Corollary
\newtheorem{lemma}{Lemma}      % Specify Lemma
\newtheorem{result}{Result}      % Specify Lemma
\newcommand{\qed}{\hfill{$\rule{6pt}{6pt}$}} %Box at end of proof
\newenvironment{proof}{\noindent{\bf Proof}:}{\qed}

\newcommand{\defeq}{\stackrel{\Delta}{=}}
\newcommand{\alice}{{\sf Alice}}
\newcommand{\bob}{{\sf Bob}}

\newcommand{\ket}[1]{| #1 \rangle}

\newcommand{\ketbra}[1]{| #1 \rangle \langle #1 |}
\newcommand{\braket}[2]{\langle #1 | #2 \rangle}

\newcommand{\Tr}{\mbox{{\rm Tr} }}
\newcommand{\trace}[1]{\left\| #1 \right\|_t}
\newcommand{\totvar}[1]{\| #1 \|_1}
\newcommand{\hech}{{\cal H}}
\newcommand{\kay}{{\cal K}}
\newcommand{\ua}{U^{A^{i-1}}}
\newcommand{\va}{V^{A^i}_{00 \rightarrow 10}}
\newcommand{\vb}{V^{B^{i-1}}_{00 \rightarrow 01}}
\newcommand{\cX}{{\cal X}}

\newcommand{\cZ}{{\cal Z}}
\newcommand{\bD}{{\mathbf D}}
\newcommand{\bd}{{\mathbf d}}
\newcommand{\ba}{{\mathbf a}}
\newcommand{\cD}{{\cal D}}
\newcommand{\cA}{{\cal A}}

\newcommand{\bx}{{\mathbf x}}

\newcommand{\bX}{{\mathbf X}}

\newcommand{\hbd}{\hat{\mathbf d}}

\newcommand{\DISJ}{{\sf{DISJ}}}

\newcommand{\protocol}{\Pi}
\newcommand{\IL}{{\mathsf{IL}}}
\newcommand{\AND}{{\sc AND }}

\newcommand{\iseq}{\stackrel{\mathrm{?}}{=}}
\newcommand{\linfinity}{{{\cal L}_{\infty}}}
\newcommand{\xor}{\oplus}
\newcommand{\embed}{{\mathsf {embed}}}
\def\E{\mathop{\rm E}}
\newcommand{\IN}{{\mathsf{IN}}}
\newcommand{\tIN}{\widetilde{\mathsf{IN}}}

\title{A lower bound for bounded round quantum communication complexity
of set disjointness}
\author{
Rahul Jain\footnotemark
\addtocounter{footnote}{-1}
\and
Jaikumar Radhakrishnan\footnotemark
\addtocounter{footnote}{-1}
\and
Pranab Sen\footnote{
School of Technology and Computer Science,
Tata Institute of Fundamental Research,
Mumbai 400005,
India.
Email:  {\sf \{rahulj, jaikumar, pranab\}@tcs.tifr.res.in}.
Rahul Jain was supported partially by the Kanwal Rekhi Career Development 
Scholarship.
}
}
\date{}

\begin{document}
\maketitle

\begin{abstract} 
We show lower bounds in the multi-party quantum communication
complexity model. In this model, there are $t$ parties where the $i$th
party has input $X_i \subseteq [n]$. These parties communicate with
each other by transmitting qubits to determine with high probability
the value of some function $F$ of their combined input
$(X_1,X_2,\ldots,X_t)$. We consider the class of functions whose
value depends only on the intersection of $X_1,X_2, \ldots,X_t$; that
is, for each $F$ in this class there is an $f_F: 2^{[n]} \rightarrow
\{0,1\}$, such that \[ F(X_1,X_2,\ldots,X_t) = f_F(X_1 \cap X_2 \cap
\ldots \cap X_t).\] 

We show that the $t$-party $k$-round communication complexity of $F$
is $\Omega(s_m(f_F)/(k^2))$, where 
$s_m(f_F)$ stands for the `monotone
sensitivity of $f_F$' and is defined by
\[ s_m(f_F) \defeq \max_{S\subseteq [n]} |\{ i: f_F(S \cup \{i\}) \neq
f_F(S)\}|.\]

For two-party quantum communication protocols for the {\em set
disjointness problem,} this implies that the two parties must exchange
$\Omega(n/k^2)$ qubits. An upper bound of $O(n/k)$ can be derived from
the $O(\sqrt{n})$ upper bound due to Aaronson and Ambainis (see also
\cite{buhrman:sqrtlog} and \cite{hoyer:sqrtlog*}).  For $k=1$, our
lower bound matches the $\Omega(n)$ lower bound observed by Buhrman
and de Wolf ~\cite{buhrman:disj} (based on a result of
Nayak~\cite{nayak:index}), and for $2\leq k \ll n^{1/4}$, improves the
lower bound of $\Omega(\sqrt{n})$ shown by Razborov~\cite{razborov:sqrt}.
(For protocols with no restrictions on the number of rounds, we can
conclude that the two parties must exchange $\Omega(n^{1/3})$ qubits.
This, however, falls short of the optimal $\Omega(\sqrt{n})$ lower
bound shown by Razborov~\cite{razborov:sqrt}.)

Our result is obtained by adapting to the quantum setting the elegant
{\em information-theoretic} arguments of Bar-Yossef, Jayram, Kumar and
Sivakumar~\cite{yossef:disj}. Using this method we can show similar lower
 bounds for the $\linfinity$ function considered in 
\cite{yossef:disj}.
\end{abstract}

\section{Introduction}

\paragraph{Classical communication complexity:} 
The communication complexity model of Yao~\cite{yao:cc} provides an
abstract setting for studying the communication required for computing
a function whose inputs are distributed between several parties.  In
its most widely studied version, there are two parties, {\alice} and
{\bob} with inputs $X_A, X_B \subseteq [n]$, who exchange messages
based on a fixed protocol in order to determine the value of some
function $F(X_A,X_B)$. The goal is to design a protocol so that the
parties need to exchange as few bits as possible. This model of
communication is relatively well-understood (see the book of
Kushilevitz and Nisan~\cite{nisan:cc}) both in the deterministic and
the randomized setting.  In this paper, we will be interested in the
randomized setting, where the parties are allowed to err with some
small probability (say at most $\frac{1}{3}$).  Tight lower bounds are
known for several functions, in this model, for example, the equality
function $X_A \iseq X_B$~\cite{yao:cc, lipton:cc}, the
set disjointness function $X_A \cap X_B \iseq
\emptyset$~\cite{kalyan:disj, razborov:omegan} and the inner-product
function $|X_A \cap X_B|\mbox{ (mod 2)}$~\cite{chor:ip}.

\paragraph{Quantum communication complexity:} The two-party quantum
communication model (see Section~\ref{subsec:quantumcomm}) was introduced by
Yao~\cite{yao:quantcc}, in order to investigate if communication costs for
computing functions distributively reduces significantly when the
parties are allowed to exchange qubits and perform quantum operations
locally. Since then, there has been a flurry of results in this
model. We will be mainly interested in the bounded error version of
this model, where the two parties are allowed to err with some small
probability (say at most $\frac{1}{3}$).  It was observed early that
for the equality and the inner-product functions the quantum model
does not provide any significant savings: the complexity of the
equality function is still $\Theta(\log n)$~\cite{kremer:quantcomm} and the
complexity of the inner-product function is still
$\Theta(n)$~\cite{kremer:quantcomm,cleve:ip}.

\paragraph{The set disjointness function:} For the set disjointness function,
however, quantum protocols were found to be strictly more powerful
than their classical randomized counterparts.  Since the communication
complexity of the set disjointness function is central to the work
presented in this paper, we describe its history in greater detail. In
the bounded error classical setting Babai, Frankl and
Simon~\cite{babai:sqrtn} showed a lower bound of
$\Omega(\sqrt{n})$. This was improved to an $\Omega(n)$ lower bound by
Kalyanasundaram and Schnitger~\cite{kalyan:disj}; their proof was
simplified by Razborov~\cite{razborov:omegan}. There is a
straightforward protocol with $n+1$ bits of communication where
{\alice} sends her entire input to {\bob}, who computes the answer and
returns it to {\alice}.  Interest in the communication complexity of
several problems related to the set disjointness function has been
revived recently because of their connection to showing lower bounds
in the classical datastream model~\cite{alon:freq, kannan:freq,
guha:freq, indyk:freq, guha:data, jayram:info,saks:freq}. One of these
problem is the $\linfinity$ promise problem: {\alice} and {\bob} are
given inputs $X_A, X_B \in \{0,1,\ldots,m\}^n$, with the promise that
either for all $i\in [n]$, $|X_A[i]-X_B[i]| \leq 1$ or there exists an
$i\in [n]$, such that $|X[i]-Y[i]| = m$; they must communicate in
order to distinguish between these two types of inputs.  For this
problem, Saks and Sun ~\cite{saks:freq} showed a lower bound of
$\Omega(n/m^2)$ in a restricted model; their lower bound was
strengthened by Bar-Yossef, Jayram, Kumar and Sivakumar~\cite{yossef:disj},
who obtained the same lower bound without any restrictions.

In the quantum setting, the set disjointness function was first addressed
by Buhrman, Cleve and Wigderson~\cite{buhrman:sqrtlog}, who showed
that there is a protocol for this problem with $O(\sqrt{n} \log n)$
bits of communication. This bound was improved to $O(\sqrt{n} c^{\log^*
n})$, where $c$ is a small constant, by Hoyer and de
Wolf~\cite{hoyer:sqrtlog*}, and recently to $O(\sqrt{n})$ by Aaronson
and Ambainis~\cite{aaronson:disj}. By a result of
Razborov~\cite{razborov:sqrt} this last bound is optimal.

\paragraph{Multi-party classical communication complexity:} There are
several ways to generalize the two-party model to the multi-party
model. In this paper, we will consider the version where there are $t$
parties $P_1,P_2,\ldots, P_t$ with respective inputs
$X_1,X_2,\ldots,X_t \subseteq [n]$. In each round of communication
some party sends a message to another party. The party who receives
the last message can determine the desired value
$F(X_1,X_2,\ldots,X_t)$ based on his current state at that
point. Recently, because of its connection to the problem of computing
{\em frequency moments} in the data stream model~\cite{alon:freq}, the
following {\em promise set disjointness} problem has been studied.  Here,
the parties are required to distinguish between two extreme types of
inputs: in the first type, $X_1,X_2,\ldots,X_t$ are pairwise disjoint;
in the second type, $X_1,X_2,\ldots,X_t$ have exactly one element in
common but are otherwise disjoint. For this problem, Chakrabarti, Khot
and Sun~\cite{chak:disj} show a lower bound of $\Omega(n/(t\log t))$,
improving an earlier $\Omega(n/t^2)$ lower bounds of Bar-Yossef,
Jayram, Kumar and Sivakumar~\cite{yossef:disj} and an $\Omega(n/t^4)$ lower
bound of Alon, Matias and Szegedy~\cite{alon:freq}.  A slight variant of
this problem, called the approximate set disjointness problem, was
considered by Nisan~\cite{nisan:disj}; the lower bounds mentioned above
apply to Nisan's version as well. The multi-party quantum
communication complexity of these problems has not been considered
before this work.

\subsection{Our results}

The upper and lower bounds on the two-party quantum communication
complexity of the set disjointness function are tight up to constant
factors, if there are no restrictions imposed on the number of rounds
(i.e. the number of messages) in the protocol. The best upper bound
uses $O(\sqrt{n})$ rounds of communication, and from it one can derive
a $k$-round protocol where the parties exchange a total of at most
$O(n/k)$ qubits. For $k=1$, Buhrman and de Wolf~\cite{buhrman:disj}
observed that the lower bound of $\Omega(n)$ follows from the results
of Nayak~\cite{nayak:index} for the index-function problem. For $k\geq
2$, Klauck, Nayak, Ta-Shma and Zuckerman~\cite{klauck:ptr} showed a
lower bound of $n^{1/k}$, but this is subsumed by
Razborov's~\cite{razborov:sqrt} lower bound of $\Omega(\sqrt{n})$
which holds even if there is no restriction on the number of
rounds. However, for small $k$, Razborov's lower bound is far from the
best upper bound known, namely $O(n/k)$. Our first result, gives lower
bounds for the two-party $k$-round communication complexity that comes
closer to the upper bound.
\begin{result}
\label{res:disj} 
The  two-party $k$-round quantum
communication complexity of the set disjointness function is
$\Omega(n/k^2)$. 
\end{result}

In fact, this lower bound holds even if the protocol is only required
to distinguish between disjoint sets and sets with exactly one element
in common. Using easy reductions one can conclude that a similar
lower bound holds for several other functions.  A function $F$ is said
to be {\em set disjointness-like} if its value depends only on the
intersection of $X_A,X_B$; that is, there is an $f_F: 2^{[n]}
\rightarrow
\{0,1\}$, such that $F(X_A,X_B) = f_F(X_A \cap X_B)$.  
We obtain a non-trivial lower on the communication complexity of such
functions $F$, if the underlying function $f_F$ has high {\em monotone
sensitivity:} $\displaystyle s_m(f_F) \defeq \max_{S\subseteq [n]} |\{
i: f_F(S \cup \{i\}) \neq f_F(S)\}|.$ \\
{\bf Result 1':} The  two-party $k$-round quantum
communication complexity of the a set disjointness-like function $F$ is 
$\Omega(s_m(f_F)/(k^2))$. 

For the $\linfinity$ promise problem we get the following.
\begin{result}
The two-party $k$-round quantum communication complexity of the
$\linfinity$ promise problem is \newline $\Omega(n/(k^3m^{(k+1)}))$.
\end{result}
We define a model for multi-party quantum communication complexity and
show the following\footnote{Our lower bound appears to contradict 
the $\tilde{O}(n/t)$ upper bound of~\cite{yossef:disj}. This is because
that upper bound is in the simultaneous message model, whereas in our 
definition of quantum protocols one is required to pass fixed length
messages from  one party to another.}.
\begin{result}
The $t$-party $k$-round quantum communication complexity of the
promise set disjointness problem is $\Omega(n/k^2)$.  [This lower bound
also holds for Nisan's approximate set disjointness problem.]
\end{result}

All our lower bounds hold even if the parties start with arbitrary
prior entanglement that is independent of the inputs.

\subsection{Techniques used}
The original lower bounds for the set disjointness problem in the
classical setting are based on deep analyses of the communication
matrix and can be said to be based on the {\em discrepancy
method}~\cite{chazelle}.  Razborov's recent $\Omega(\sqrt{n})$
lower bound for quantum protocols also uses the discrepancy method.
The discrepancy method for quantum protocols was formulated explicitly
by Kremer~\cite{kremer:quantcomm} (see also Klauck~\cite{klauck:qcc} and
Yao~\cite{yao:quantcc}), but Razborov's proof extends it substantially by
developing interesting and powerful tools based on the spectral theory
of matrices.

Recently, however, Bar-Yossef et al.~\cite{yossef:disj} proposed an
information-theoretic approach for studying set disjointness-like problems
in the classical setting. Using a refinement of the notion of
information of communication protocol originally defined by
Chakrabarti, Shi, Wirth and Yao~\cite{chak:direct}, they showed that a linear
lower bound for the set disjointness problem follows from $\Omega(1)$
lower bound on a certain information cost of a two-party communication
protocol computing the \AND of just two bits!  Their work provided a
compelling and beautiful illustration of information-theoretic tools
in the analysis of communication protocols.

We adapt their approach to the quantum setting. In order to bring out
the contribution of this paper more clearly, we will now informally
describe the information-theoretic argument underlying their proof and
discuss how we adapt them to the quantum setting. The argument has two
parts: in the first part, using a direct-sum argument for information
from Bar-Yossef et al.~\cite{yossef:disj}, one reduces the set
disjointness problem to a communication problem associated with the
\AND of two bits (one with {\alice} and one with {\bob}); in the second
part, one shows that this problem on two bits is hard.

\paragraph{The information cost approach:} The first part of the
argument is based on the notion of information cost of communication
protocols, defined (by~\cite{chak:direct}) to be the mutual information between
the inputs (which are assumed to come from some distribution) and the
transcript of the protocol.  Bar-Yossef et al.~\cite{yossef:disj} examine the
information cost of the protocol for several distributions. Let the
number of bits transmitted by the protocol be $c$. Then, the
information cost is also bounded by $c$ for each distribution.

At this point it will be convenient to view the inputs of {\alice} and
{\bob} as elements of $\{0,1\}$ and the set disjointness function as
$\bigvee_{i=1}^n X_A[i] \wedge X_B[i]$. A typical distribution
considered by Bar-Yossef et al. is defined as follows. For each $i$,
independently, one party is given the input $0$ and the other
party is given a random bit. Using the sub-additivity property of
mutual information, one concludes that the sum over $i$ of the 
mutual information between the transcript and the input $X_A[i]$ is
bounded by $c$; a similar statement holds for {\bob}'s inputs. It is
then not hard to argue using a standard averaging argument that there
is an $i$ and a product distribution $D^*$, for inputs $(X_A[j], X_B[j] : j
\neq i)$ such that the following conditions hold:
\begin{itemize}
\item For all $j\neq i$, $X_A[j] \wedge X_B[j]=0$ (with probability 1).
\item If $X_A[i]$ is set to zero and $X_B[i]$ is chosen at random 
(and the remaining bits are chosen according to the product
distribution $D^*$), then the mutual information between the
transcript and $X_B[i]$ is at most $2c/n$; similarly, if 
$X_B[i]$ is set to 0 and $X_A[i]$ is chosen at random 
(an the remaining bits are chosen according to the product
distribution $D^*$), then the mutual information between the
transcript and $X_A[i]$ is at most $2c/n$. 
\end{itemize}
From the first condition, by viewing $(X_A[j], X_B[j] : j \neq i)$ as
private random bits of the two-parties, we obtain from the protocol
for set disjointness a protocol that computes the \AND of the two bits
$X_A[i]$ and $X_B[i]$. The stage is thus set for analysing the
information cost of computing the \AND function: a lower bound of
$\epsilon$ on this quantity translates to a lower bound of
$\Omega(\epsilon n)$ on the communication complexity of the 
set disjointness function. 

In order to implement this programme in the quantum setting, one has
to define a notion of information cost for quantum protocols. It is
not immediately clear how this can be done, because quantum operations
are notorious for destroying the states on which they act; in
particular, it is not reasonable to expect that the complete
transcript of all messages is part of the final global state of the
algorithm. Even if the complete transcript is available in the final
global state of the algorithm, it may not contain any information
about the inputs of either party. If the parties are allowed prior
entanglement, then using quantum teleportation, one can implement any
protocol such that the messages are classical and completely
random. So, the transcript will just be a random string of length $c$
independent of the actual inputs!

\paragraph{The definition of information loss for quantum protocols:} 
We address these difficulties by considering the information carried
by each message separately. As observed above messages may themselves
carry no information, so we examine the information carried in the
message by including the context in which it is received. For example,
consider a protocol for the \AND problem. Fix some distribution for
the inputs of {\alice} and {\bob}.  We account for the information
carried in a message sent by {\alice} to {\bob}, by considering the
mutual information between {\alice}'s input and the {\em entire} state
of {\bob}, including the message just received. The information loss
(we use the term loss instead of cost) of the protocol (for the given
distribution) is defined to be the sum of these quantities (both for
{\alice} and {\bob}) taken over all rounds. With this definition, the
arguments of~\cite{yossef:disj} are easily carried over to the quantum
setting. We can then conclude that if the information loss of
computing the \AND of two bits is $\epsilon$ then the communication
complexity of the set disjointness function is $\Omega(n\epsilon)$.

We have arrived at the second part of the programme, that is, to
show non-trivial lower bounds on the information loss of computing the
\AND of two bits. In the original argument of ~\cite{yossef:disj} this was
achieved by a direct argument using certain distance measures between
probability distributions. Since, we are working with a different
notion of information loss, this argument does not appear to be
immediately applicable in our case; so, instead of reviewing it, we
will now directly describe our argument. We are given a quantum
communication protocol for computing the \AND function. We consider
two kinds of inputs: first, {\alice} has 0 and {\bob} has a random
bit; second, {\bob} has a 0 and {\alice} has a random bit. Suppose we
are given that for such distributions at no stage does a receiver of a
message gain more than $\epsilon$ bits of information about the input
of the sender. We wish to show that if $\epsilon$ is very small, then
this leads to a contradiction. Our argument can be understood at an
intuitive level in the framework of round-elimination. Suppose, Alice
sends the first message. We know that her message does not deliver
much information about her input to {\bob}, that is, the combined
state of {\bob} at the end of the first round is essentially the same
when the {\alice}'s input is $0$ and when {\alice}'s input is $1$. So,
{\alice} might as well send exactly the same message in the two cases,
and incur a small error in the correctness. That is, no matter what
her actual input is, {\alice} sends her first message assuming that
her input is 0. Since we allow prior entanglement, we can eliminate
this round of Alice, and obtain a protocol with one fewer round of
communication. Now, it is {\bob}'s turn. Our hypothesis says that his
second message does not deliver much information about his input to
{\alice}, when her input is $0$. But the modified protocol so far has
proceeded as if {\alice}'s input is $0$ (even though her actual input
might be something else). We can thus eliminate Bob's first message as
well. If $\epsilon$ is small, then the increase in error probability
on account of this manoeuvre is also small.  Proceeding in this manner
we eliminate all rounds. But it is obvious that if the parties
exchange no messages they cannot compute any non-trivial function
unless one allows huge error probability. Since, there are at most $k$
rounds of communication, this gives us a lower bound of the form
$\epsilon \geq \epsilon(k)$. Using these ideas one can show an
$\Omega(n/k^2)$ lower bound on two-party quantum communication
complexity of the set disjointness function.

There are two aspects of our proof that require further comment. 

\paragraph{Local transition:} Recall the argument used above to
eliminate {\alice}'s first message. We know that {\bob}'s state is
roughly the same even if {\alice} generates her message assuming that
her input is $0$. However, this does not immediately imply that the
error probability of the protocol is not changed much. The final
answer is not just a function of {\bob}'s state but the combined state
of {\alice} and {\bob}. In particular, even though the {\bob}'s state
is similar after the first round for the two inputs of {\alice}, his
work qubits might be entangled with {\alice}'s qubits differently in the
two cases. This problem arises often in round elimination arguments
and by now standard solutions exist for it by considering the {\em
fidelity}~\cite{jozsa:fidelity} between quantum states. This allows
{\alice} to perform a {\em local transition}~\cite{klauck:ptr} on her
work qubits, in order to restore them to the correct state should she
discover later that her actual input is different from what was
assumed while generating her first message to {\bob}.

\paragraph{A paradox?:} In our notion of information loss of quantum
protocols it is important that the parties start in a {\em pure}
global state. In fact, this notion is unsuited for classical
randomized communication complexity. Consider the following classical
protocol for computing the \AND of two bits $(a,b)$. {\alice} sends
{\bob} a random bit $r$, retaining a copy of $r$ if and only if
$a=1$. {\bob} sends {\alice} $r\xor b$; if $a=1$, {\alice} can recover
$b$ using the copy of $r$ she has and determine $a\wedge b$.  Now,
clearly, the first message does not deliver any information to
{\bob}. Furthermore, when {\alice} has a $0$, {\bob}'s message
delivers no information about his inputs, because {\alice} does not
retain a copy of $r$ in this case. So, according to our definition
this protocol has zero information loss for both the distributions
considered above. Yet, the protocol computes the \AND correctly!
Interestingly, no such quantum protocol starting with a pure global
state is possible.

\subsection{The rest of the paper}
In the next section, we give some the definition and notation used in
the rest of the paper.  In Section~\ref{sec:lowerbound}, we prove
Result~\ref{res:disj}. Result~2 and Result~3 also follow using
similar arguments, but their proofs are not included in this abstract. 

\section{Preliminaries}

\subsection{Quantum communication }
\label{subsec:quantumcomm}
We define $t$-party quantum communication protocols which are a
natural extension of the two-party quantum communication protocols as
defined by Yao~\cite{yao:quantcc}.  Let $f: \cX_1 \times \cX_2 \cdots
\cX_t \rightarrow \cZ$ be a function.  There are $t$ parties, $P_1,
P_2, \cdots, P_t$, who hold qubits.  When the communication protocol
$\protocol$ starts, $P_i$ holds $\ket{x_i}$ where $x_i \in \cX_i$
together with some ancilla qubits in the state $\ket{0}$.  These
parties may also share an input independent prior entanglement (say
$\ket{\psi}$).  Different parties possess different qubits of
$\ket{\psi}$.  The parties take turns to communicate to compute
$f(x_1,x_2, \cdots, x_t )$.  Suppose it is $P_1$'s turn to communicate
to $P_2$.  $P_1$ can make an arbitrary unitary transformation on her
qubits and then send one or more qubits to $P_2$. The number of qubits send
is predetermined and is independent of the input $x_1$. Sending qubits does not
change the overall superposition, but rather changes the ownership of
the qubits, allowing $P_2$ to apply her next unitary transformation on
her original qubits plus the newly received qubits. At the end of the
protocol, the last recipient of a message performs a von Neumann
measurement in the computational basis of some qubits in her
possession (the `answer qubits') to output an answer $\protocol(x_1,
x_2, \cdots, x_t)$.  We say that protocol $\protocol$ computes $f$ with
$\delta$-error in the worst case (or simply with error $\delta$), if
$\max_{x_1, x_2, \cdots, x_t} \Pr[\protocol(x_1, x_2, \cdots, x_t) 
\neq f(x_1, x_2, \cdots, x_t)] \leq \delta$.  The
communication cost of $\protocol$ is the number of qubits exchanged in
$\protocol$ between all the parties.  The $k$-round $\delta$-error
quantum communication complexity of $f$, denoted by $Q^k_\delta(f)$,
is the communication cost of the best $k$-round $\delta$-error quantum
protocol with prior entanglement for $f$. When $\delta$ is omitted, we
mean that $\delta=\frac{1}{3}$.

We require that the parties make a `safe' copy of their inputs (using,
for example, CNOT gates) before beginning protocol $\protocol$.  This
is possible without loss of generality because the inputs are in
computational basis states.  Thus, the input qubits of the parties are
never sent as messages, their state remains unchanged throughout the
execution of $\protocol$, and they are never measured i.e. some work
qubits are measured to determine the result $\protocol(x_1, x_2,
\cdots, x_t)$. We call such protocols {\em safe}, and henceforth, we
will assume that all our protocols are safe.

Suppose $A, B, C$ are three disjoint finite dimensional
quantum systems having some joint density matrix $\rho$. 
Let $\rho_A$ be the reduced density matrix of A. Then
$S(A) \defeq S(\rho) \defeq -\Tr \rho \log \rho$ is the
{\em von Neumann entropy} of $A$.  The 
{\em mutual information} of $A$ and $B$ is defined as 
$I(A : B) \defeq S(A) + S(B) - S(AB)$. The 
{\em conditional mutual information} of $A$ and $B$ given $C$ is
defined as 
$I((A : B) \mid C) \defeq S(AC) + S(BC) - S(C) - S(A B C)$.
If $C$ is a classical random variable taking the classical
value $\ket{c}$ with probability $p_c$, it
is easy to see that
$I((A : B) \mid C) = \sum_c p_c I(A^c : B^c)$, where $(AB)^c$ denotes
the joint density matrix of $A$ and $B$ when $C = \ket{c}$. We also
write $I(A:B \mid C=c)$ for $I(A^c:B^c)$.

\begin{fact}[see \cite{cleve:ip}] 
\label{fact:cleve}
Let {\alice} have a classical random variable $X$.  Suppose {\alice}
and {\bob} share a pure state on some qubits (a prior entanglement)
independent of $X$. Initially {\bob}'s qubits have no information about
$X$.  Now let {\alice} and {\bob} run a quantum communication protocol,
at the end of which {\bob}'s qubits possess $m$ bits of information
about $X$. Then, {\alice} has to totally send at least $m/2$ qubits to
{\bob}.
\end{fact}
 
\begin{fact}[Sub-additivity of information,see ~\cite{klauck:ptr}]
\label{fact:subadditivity}
Let $D$ be a classical random variable.
Let $X_1, \ldots, X_n$ be classical random variables which are
independent given $D$.
Let $M$ be a quantum encoding of $X \defeq X_1 \ldots X_n$.
Then, $I((X:M)\mid D) \geq \sum_{i=1}^n I((X_i : M)\mid D)$.
%Also, if $M$ is $n$ qubits long, then $I(X:M) \leq n$.
\end{fact}

\begin{definition}[Trace distance]
Let $\rho, \sigma$ be density matrices in the same finite dimensional
Hilbert space. 
The trace distance
between $\rho$ and $\sigma$ is defined as follows: 
$\trace{\rho - \sigma} \defeq 
 \Tr \sqrt{(\rho - \sigma)^{\dagger} (\rho - \sigma)}$. 
\end{definition}

\begin{definition}[Fidelity]
Let $\rho$, $\sigma$ be density matrices in the same finite
dimensional Hilbert space $\hech$. Their fidelity is defined as
$B(\rho, \sigma) \defeq \sup_{\kay, \ket{\psi}, \ket{\phi}}
|\braket{\psi}{\phi}|, $ where $\kay$ ranges over all finite
dimensional Hilbert spaces and $\ket{\psi}, \ket{\phi}$ range over all
purifications of $\rho, \sigma$ respectively in $\hech \otimes \kay$.
\end{definition}

\begin{fact}[see~\cite{aharonov:mixed}]
\label{fact:totvartrace}
Let $\rho, \sigma$ be density matrices in the same finite
dimensional Hilbert space $\hech$. Let ${\cal F}$ be a
measurement (POVM) on $\hech$. Then, 
$\totvar{{\cal F} \rho - {\cal F} \sigma} \leq \trace{\rho - \sigma}$.
\end{fact}

The following lemmas are derived in the appendix.
\begin{lemma}
\label{lem:localtransition}
Let $\rho_1, \rho_2$ be two density matrices in the same finite
dimensional Hilbert
space $\hech$, $\kay$ any Hilbert space of dimension at least
the dimension of $\hech$, and $\ket{\phi_i}$ any purifications
of $\rho_i$ in $\hech \otimes \kay$. Then, there is a local
unitary transformation $U$ on ${\cal K}$ that maps $\ket{\phi_2}$
to $\ket{\phi_2'} \defeq (I \otimes U) \ket{\phi_2}$ ($I$ is the
identity operator on $\hech$) such that
\begin{displaymath}
\trace{\ketbra{\phi_1} - \ketbra{\phi_2'}} 
        \leq 2\sqrt{1 - B(\rho_1, \rho_2)^2} 
        \leq 2\sqrt{2(1 - B(\rho_1, \rho_2))}. 
\end{displaymath}
\end{lemma}

\begin{lemma}
\label{lem:linquantum}
Suppose $X$ and $Q$ are disjoint quantum systems, where $X$ is a classical 
random variable uniformly distributed over $\{0,1\}$ and $Q$ is a quantum 
encoding $x \rightarrow \sigma_x$ of $X$. Then, 
\(1 - B(\sigma_1, \sigma_2) \leq  I(X:Q). \)
\end{lemma}

\subsection{Conditional information loss}

Let $D$, $X_A$ and $X_B$ be random variables taking values in some
finite sets $\cD$, $\cX_A$ and $\cX_B$ respectively. We say that $D$
partitions $X=(X_A,X_B)$ if for all $d\in \cD$, $X_A$ and $X_B$ are 
independent conditioned on the event $D=d$.
Given random variables $X$ and $D$, the random variable
$(X,D)^n$ is obtained by taking $n$ independent copies of $(X,D)$.  Thus,
$(X,D)^n$ takes values in $(\cX_A \times
\cX_B)^n$ which we identify with $\cX_A^n \times \cX_B^n$.
Suppose $D$ partitions $X$, and $(\bX,\bD)=(X,D)^n$, then it is easy
to verify that $\bD$ partitions $\bX$.

\begin{definition}[Embedding]
\label{def:embedding}
For ${\ba} \in \cA^n$, $j \in [n]$, and $u \in \cA$, let $\embed({\ba},
j, u)$ be the element of $\cA^n$ obtained by replacing $\ba[j]$ by $u$,
that is, $\embed(\ba, j, u)[i] \defeq \ba[i]$ for $i \neq j$, and
$\embed(\ba, j, u)[j]\defeq u$.
\end{definition}

\begin{definition}[Collapsing input]
\label{def:collapsing}
Suppose $F: \cX^n \rightarrow \cZ$. We say that $\bx \in X^n$ 
collapses $F$ to the function $h: \cX \rightarrow \cZ$ if for all $u 
\in \cX$, $F(\embed(\bx, j, u)) = h(u)$.
We say that a random variable $X$ taking values in $\cX^n$ 
collapses $F$ to $h$ if it is collapses $F$ to $h$ with probability
$1$. 
\end{definition}

\begin{definition}[Conditional Information loss] \label{defn:il}
Let $\protocol$ be a two-party $k$-round $\delta$-error quantum
protocol for computing $F:
\cX_A \times \cX_B \rightarrow \cZ$.  Let {\alice} start the protocol
and let $A^i B^i$ be the joint state of {\alice} and {\bob} just after
the $i$th message has been received. Let $X=(X_A,X_B)$ be random
variable taking values in $\cX_A \times \cX_B$ which is partitioned by
the random variable $D$.  Then, the {\em conditional information loss}
of $\protocol$ under $(X, D)$ is defined by
\[ 
\textstyle
\IL(\Pi \mid (X,D)) \defeq \sum_{i=1, ~i~\mbox{{\rm odd}}}^k
I(X_A:B^i \mid D) + \sum_{i=1, ~i~\mbox{{\rm even}}}^k I(X_B:A^i \mid
D).\] 
The $k$-round $\delta$-error conditional information loss of $F$
under $(X, D)$, denoted by $\IL_{k,\delta}(F \mid (X,D))$, is the
minimum $\IL(\Pi \mid (X,D))$ taken over all $k$-round $\delta$-error
quantum protocols $\Pi$ for $F$. [Note that $\delta$ bounds the error
for {\em all} inputs. In particular, this error bound applies even to
inputs not in the support of $X$.]
\end{definition}

\section{Lower bound for set disjointness}
\label{sec:lowerbound}
\begin{lemma} 
\label{lem:disjtoand}
Let $F: \cX_A^n \times \cX_B^n \rightarrow \cZ$. Let $X$ be a random
variable taking values in $\cX \defeq \cX_A \times \cX_B$; suppose $X$
is partitioned by a random variable $D$ taking values in some set
$\cD$.  Let $(\bX,\bD) = (X,D)^n$. Suppose $\bX$ collapses $F$ to the
function $h:\cX_A \times \cX_B \rightarrow \cZ$.  Then,
\(\IL_{k,\delta}(h \mid (X,D)) \leq \frac{2k}{n} Q^k_\delta(F).\)
\end{lemma}
\begin{proof}
Suppose $\protocol$ is a $k$-round $\delta$-error quantum protocol 
for $F$ with total communication $c$. Let us assume that {\alice}
starts the communication. Our goal is to show that there is a
$k$-round $\delta$-error protocol for $h$ with small information loss
under $(X,D)$. While analysing $\protocol$, we will need to maintain
that the combined state of {\alice} and {\bob} is pure at all
times. However, we will run $\protocol$ on random inputs drawn from
certain product distributions. In such a situation, we will adopt the
following convention. We will assume that in addition to the usual
input registers $\IN_A$, {\alice} has another set of registers
$\tIN_A$. When we require that {\alice}'s inputs be some random
variable $X_A$, we in fact, start with the following state in the
registers $\IN_A{\tIN_A}$:
\(\sum_{x\in \cX_A} \sqrt{p_x} \ket{x}\ket{x},\)
where $p_x\defeq \Pr[X_A=x]$.  Similarly, we simulate {\bob}'s random
input $X_B$ in registers $\IN_B$ and $\tIN_B$. Then, we run the
protocol $\Pi$ as before with input registers $\IN_A$ and
$\IN_B$. During this execution no quantum gates are applied to
registers $\tIN_A$ and $\tIN_B$. From now on $\bX_A$ (similarly
$\bX_B$) denotes the state of the registers $\IN_A$, which stays
constant because the protocol $\Pi$ is safe. In this revised protocol
$\protocol'$, let $A^iB^i$ denote the state of the entire system
immediately after the $i$th message has been received (note that $A^i$
includes the register $\tIN_A$ and $B^1$ includes $\tIN_B$). Consider
the execution of $\protocol'$ on input $\bX=(\bX_A,\bX_B)$ conditioned
on $\bD=\bd$; note that under this condition $\bX_A$ and $\bX_B$ are
independent and the convention described above for simulating random
inputs applies.  Then, we have
\[
\textstyle
\forall i, 1 \leq i \leq k, i ~ \mbox{odd}, ~~ 
\sum_{j=1}^n I((\bX_A[j] : B^i) \mid \bD=\bd) \leq I((\bX_A : B^i) \mid \bD=\bd)
\leq 2c.\]
The first inequality above follows from Fact~\ref{fact:subadditivity}
because by our definition of $(\bX,\bD)$, $(\bX_A[j]: 1\leq j \leq n)$
are independent random variables when conditioned on $\bD=\bd$; the second
inequality follows from Fact~\ref{fact:cleve}. 

Averaging over the possibilities for $\bD$, we obtain:
\(
\textstyle
\forall i, 1 \leq i \leq k, i ~ \mbox{odd}, ~~ 
\sum_{j=1}^n I((\bX_A[j] : B^i) \mid \bD) \leq 2c.
\)
Similarly, we obtain
\(
\textstyle
\forall i, 1 \leq i \leq k, i ~ \mbox{even}, ~~  
\sum_{j=1}^n I((\bX_B[j] : A^i) \mid \bD) \leq  2c.  
\) 
Summing these inequalities over all rounds $i$, we obtain
\[
\textstyle
\sum_{j=1}^n \left(
\sum_{i=1, ~i~\mbox{{\rm odd}}}^k  I(\bX_A[j] : B^i \mid \bD) +   
\sum_{i=1, ~i~\mbox{{\rm even}}}^k I(\bX_B[j] : A^i \mid \bD)
\right) \leq 2ck, 
\]
which implies:
\begin{equation}
\textstyle
\exists j, 1 \leq j \leq n, 
\sum_{i=1, ~i~\mbox{{\rm odd}}}^k  I((\bX_A[j] : B^i) \mid \bD) +   
\sum_{i=1, ~i~\mbox{{\rm even}}}^k I((\bX_B[j] : A^i) \mid \bD) 
\leq \frac{2ck}{n}. \label{eq:fixedj}
\end{equation}
Fix a value of $j$ so that the last inequality holds.
For $\bd \in \cD^n$, let
\begin{equation}
\textstyle
I(\bd) \defeq \sum_{i=1, ~i~\mbox{{\rm odd}}}^k  I((\bX_A[j] : B^i)
\mid \bD=\bd) +  \sum_{i=1, ~i~\mbox{{\rm even}}}^k I((\bX_B[j] : A^i) \mid
\bD=\bd). \label{eq:defI}
\end{equation}
Then, from (\ref{eq:fixedj}), and the definition of conditional mutual
information $\E_\bD[I(\bD)] \leq \frac{2ck}{n}$. 

We will now obtain a protocol for $h$ by `embedding' its input as the
$j$th input of $\protocol'$. Using a straightforward averaging
argument we first fix a value $\hbd \in \cD^n$ so that
\begin{equation}
\sum_{d \in \cD} \Pr[D=d] I(\embed(\hbd,j,d))=\E_D[I(\embed(\hbd,j,D)) \leq 
\frac{2ck}{n}. \label{eq:hd}
\end{equation}
Consider the following protocol $\protocol_h$ for computing
$h(u_A,u_B)$. On input $u_A \in \cX_A$, {\alice} prepares her input
registers as follows. In the registers $(\IN_A[\ell],\tIN_A[\ell]:
\ell \neq j)$ {\alice} places the superposition $\sum_{x \in
\cX^{n-1}} \sqrt{p_x}
\ket{x}\ket{x}$, where $p_x= \Pr[ (\bX_A[\ell]: \ell \neq j) = x \mid
\bD=\hbd]$; register $\IN_A[j]$ is set to
$\ket{u_A}$. On input $u_B \in \cX_B$, {\bob} prepares his
input registers in a similar fashion. Then, {\alice} and {\bob} apply
the protocol $\protocol'$, treating $\IN_A$ and $\IN_B$ as input
registers. Note that $\tIN_A$ and $\tIN_B$ do not exist in $\Pi_h$.

We need to verify that this protocol for computing $h$ has two
properties. First, it computes $h$ correctly with high
probability. For this, we note that in this protocol, at all times,
the state of the registers that were present in the original protocol
$\protocol$ (that is all registers except $\tIN_A$ and $\tIN_B$) is
identical to their state when the original protocol $\protocol$ is run with
input $\embed(\bX,j,(u_A,u_B))$ conditioned on the event $\bD=\hbd$. 
Since $\bX$ collapses $F$ to $h$, we conclude that $\protocol_h$ computes
$h(u_A,u_B)$ with probability at least $1-\delta$.

Second, we need verify that $\IL(\Pi_h \mid (X,D))$ is small.  We
expand the LHS of (\ref{eq:hd}) using the definition (\ref{eq:defI})
of $I(\bd)$ and show that each term in it is at least the corresponding
term in $\IL(\protocol_h \mid (X,D))$.  For example, consider the term $I(X_A
: B^i \mid (D=d))$ in the definition of $\IL(\protocol_h \mid (X,D))$.
Note that the state $(X_A,B^i)$ of $\protocol_h$ on input $X$
conditioned on $D=d$, is identical to the state obtained from
$(\bX_A[j],B^i)$ of $\protocol'$ by omitting the register $\tIN_B[j]$,
when $\protocol'$ is run on input $\bX$ conditioned on
$\bD=\embed(\hbd,j,d)$.  It follows from the monotonicity property of
information that $I(X_A : B^i \mid (D=d))$ is at most $I(\bX_A[j] :
B^i \mid (\bD=\embed(\hbd,j,d)))$. We can then conclude (details
omitted) that
\( \IL(\protocol_h, (X,D)) \leq \frac{2ck}{n}.\)
\end{proof}

As in \cite{yossef:disj}, let $D$ be a random variable taking values
in $\{A,B\}$ uniformly. Let $\cX_A, \cX_B=\{0,1\}$ and $X=(X_A,X_B)$
be a random variable taking values in $\cX_A \times \cX_B =
\{0,1\}^2$, whose correlation with $D$ is described
\( \Pr[X=00\mid D=A], \Pr[X=10 \mid D=A], \Pr[X=00\mid D=B],\Pr[X=01
\mid D=B]=\frac{1}{2}.\)
It is clear that conditioned on $D=A$ and $D=B$, 
$X_A$ and $X_B$ are independent. Note that $X^n$ collapses {\DISJ} to
{\AND}. We now show a lower bound for the conditional information loss
of {\AND} under $(X,D)$. 
\begin{lemma}
\label{lem:lowerboundand}
Let $(X, D)$ be as above. Let $\epsilon > 0$. Then 
$\IL_{k,\delta}(\mbox{\AND} \mid (X,D)) \geq 
\frac{(1 - 2\epsilon)^2}{4k}.$
\end{lemma}
\begin{proof}
Let $\protocol$ be a $k$-round $\epsilon$-error quantum protocol for
\AND with $\eta \defeq \IL(\protocol, (X,D)) =\IL_{k,\delta}(\mbox{\AND}
\mid (X,D))$.  Consider the situation in $\protocol$ just after the
$i$th message has been sent. Let $m^i$ denote the qubits of the $i$th
message. Let $X_A, X_B$ denote the random variables corresponding to
{\alice}'s and {\bob}'s inputs respectively in $\protocol$.  Suppose
$(X_A, X_B) = (x, y)$.  Let $\ket{\phi^i_{xy}}$ be the global state vector
of {\alice}'s and {\bob}'s qubits, and let $A^i, B^i$ denote
{\alice}'s qubits and {\bob}'s qubits respectively at this point in
time.  Suppose $m^i$ is sent from {\alice} to {\bob}. Then $B^i =
B^{i-1} \cup m^i$ and $A^{i-1} = A^i \cup m^i$, and the unions are
over disjoint sets of qubits.  Let $\alpha^i_{xy},
\beta^i_{xy}$ denote the reduced density matrices of $A^i, B^i$ in the
state $\ket{\phi^i_{xy}}$.  Define $c_B^i \defeq I(X : B^i \mid D)$
and $c_A^i \defeq I(Y : A^i \mid D)$.  Let $\ua$ denote the
unitary transformation that {\alice} applies to $A^{i-1}$ after receiving
the $(i-1)$st message from {\bob}, in order to prepare the $i$th
message. Then, $\ket{\phi^i_{xy}} = \ua \ket{\phi^{i-1}_{xy}}$. 
By Lemma~\ref{lem:linquantum}, 
$1 - B(\beta^i_{00},\beta^i_{10}) \leq 
I((X : B^i) \mid Y = 0) \leq
2 c^i_B$ and       
$
1 - B(\alpha^{i-1}_{00},\alpha^{i-1}_{01}) \leq 
I((Y : A^{i-1}) \mid X = 0) \leq
2 c_A^{i-1}
$.
To keep our notation concise, for state vectors $\ket{\phi}$ and
$\ket{\psi}$ we write $\trace{\ket{\phi} - \ket{\psi}}$ instead of 
$\trace{\ketbra{\phi} - \ketbra{\psi}}$.
By Lemma~\ref{lem:localtransition}, there exist unitary transformations
$\vb$ acting on $B^{i-1}$ and $\va$ acting on $A^i$ such that
\begin{equation}
\label{eq:bandainfo}
\trace{\vb \ket{\phi^{i-1}_{00}} - \ket{\phi^{i-1}_{01}}} \leq
4 (c_A^{i-1})^{1/2}
~~\mathrm{and} ~~
\trace{\va \ket{\phi^i}_{00} - \ket{\phi^i_{10}}} \leq
4 (c_B^i)^{1/2}.
\end{equation}

Define
$\delta_{i-1} \defeq 
\trace{\vb \ket{\phi^{i-1}_{10}} - \ket{\phi^{i-1}_{11}}}$.
Using the unitary invariance and triangle inequality of the trace norm,
the fact that unitary transformations on disjoint sets of qubits
commute, and (\ref{eq:bandainfo}),
\begin{eqnarray*}
\delta_i 
&   =  & \trace{\va \ket{\phi^i_{01}} - \ket{\phi^i_{11}}} 
   =   \trace{(\vb)^{-1} \va \ket{\phi^i_{01}} 
                - (\vb)^{-1} \ket{\phi^i_{11}})} \\ 
&   =  & \trace{\va (\vb)^{-1} \ket{\phi^i_{01}} 
                - (\vb)^{-1} \ket{\phi^i_{11}}}  \\
&   =  & \trace{\va (\vb)^{-1} \ua \ket{\phi^{i-1}_{01}} 
                - (\vb)^{-1} \ua \ket{\phi^{i-1}_{11}}} \\  
&   =  & \trace{\va \ua (\vb)^{-1} \ket{\phi^{i-1}_{01}} 
                - \ua (\vb)^{-1} \ket{\phi^{i-1}_{11}}} \\  
& \leq & \trace{\va \ua (\vb)^{-1} \ket{\phi^{i-1}_{01}} 
                - \va \ua \ket{\phi^{i-1}_{00}}} + \\
&      & \trace{\va \ua \ket{\phi^{i-1}_{00}} 
                - \ua \ket{\phi^{i-1}_{10}}} 
         + \trace{\ua \ket{\phi^{i-1}_{10}} 
                  - \ua (\vb)^{-1} \ket{\phi^{i-1}_{11}}} \\  
&   =  & \trace{(\vb)^{-1} \ket{\phi^{i-1}_{01}} 
                - \ket{\phi^{i-1}_{00}}} +  
         \trace{\va \ket{\phi^i_{00}}
                - \ket{\phi^i_{10}}} + 
         \trace{\ket{\phi^{i-1}_{10}} 
                - (\vb)^{-1} \ket{\phi^{i-1}_{11}}} \\ 
&   =  & \trace{\ket{\phi^{i-1}_{01}} - \vb \ket{\phi^{i-1}_{00}}} 
         + \trace{\va \ket{\phi^i_{00}} - \ket{\phi^i_{10}}} 
         + \trace{\vb \ket{\phi^{i-1}_{10}} - \ket{\phi^{i-1}_{11}}} \\ 
& \leq &  4 (c_A^{i-1})^{1/2} + 4 (c_B^i)^{1/2} 
          + \delta_{i-1}.
\end{eqnarray*}

It is easy to check that $\delta_0 = 0$. Hence using concavity of the
fourth root function, $\delta_k \leq 4k
\left(\frac{\eta}{k}\right)^{1/2}$. Now a correct $k$-round
$\epsilon$-error protocol for \AND must have (from
Fact~\ref{fact:totvartrace} and using the fact that a local unitary
transformation does not affect the density matrix of the remote
system), $\delta_k \geq \trace{\beta^k_{10} - \beta^k_{11}}\geq 2 -
4\epsilon$. Hence, $\eta \geq \frac{(1-2 \epsilon)^2}{4k}$.
\end{proof}

The following is now immediate from Lemma~\ref{lem:disjtoand} and 
Lemma~\ref{lem:lowerboundand}.
\begin{theorem}
\label{thm:main}
Any two-party $k$-round bounded error quantum protocol for the set
disjointness problem needs to have communication cost at least
$\Omega\left(\frac{n}{k^2}\right)$.
\end{theorem}

\begin{corollary}
Any two-party bounded error quantum protocol for the set disjointness
problem needs to have communication cost at least $\Omega\left(n^{1/3}
\right)$.
\end{corollary}

\subsection*{Acknowledgements}
Our original proof gave a lower bound of $\Omega(n/k^4)$ for set
disjointness. We later improved it to $\Omega(n/k^2)$ using an
inequality from \cite{lin}. Hartmut Klauck independently pointed out
to us that similar improvements can also be obtained using an
inequality from~\cite{dacunha:BleqS}. We thank him for sharing with us
his insights and pointing reference~\cite{dacunha:BleqS} to us.

\bibliography{disj}

\appendix 
\section{Quantum information theory background}
In this section we give some basic quantum information-theoretic
definitions and facts which will be
useful in stating and proving our main results.  For an excellent
introduction to quantum information theory, see the book by Nielsen
and Chuang~\cite{nielsen:quant}.

Suppose $P, Q$ are probability distributions on the same finite
sample space $[k]$. Their {\em total variation distance} is
defined as follows:
$\totvar{P - Q} \defeq \sum_{i \in [k]} |P(i) - Q(i)|$. The
quantum generalisation of the total variation distance of a pair
of probability distributions is the {\em trace distance} of a 
pair of density matrices. Recall that a density matrix over a 
finite dimensional Hilbert space $\hech$ is a unit trace, Hermitian,
positive semidefinite linear operator on $\hech$.

\begin{definition}[Relative Entropy]
If $\rho, \sigma$ are density matrices in the same Hilbert space,
their  relative entropy is defined as $S(\rho \| \sigma) \defeq
\Tr (\rho (\log \rho - \log \sigma))$.
\end{definition}

Let $\rho$ be a density matrix in a finite dimensional Hilbert space
$\hech$. Suppose ${\cal F}$ is a measurement (POVM) on $\hech$.
Then ${\cal F} \rho$ denotes the probability distribution on the 
(finite number of) possible outcomes of ${\cal F}$ got by performing
the measurement ${\cal F}$ on the state $\rho$.
The following fundamental facts (see~\cite{aharonov:mixed}) show that
both the trace distance and relative entropy only decrease on performing
a measurement.
\begin{fact}
\label{factappen:totvartrace}
Let $\rho, \sigma$ be density matrices in the same finite
dimensional Hilbert space $\hech$. Let ${\cal F}$ be a
measurement (POVM) on $\hech$. Then, 
$\totvar{{\cal F} \rho - {\cal F} \sigma} \leq \trace{\rho - \sigma}$.
\end{fact}

\begin{fact}
\label{fact:monorelentropy}
Let $\rho, \sigma$ be density matrices in the same finite
dimensional Hilbert space $\hech$. Let ${\cal F}$ be a
measurement (POVM) on $\hech$. Then, 
$S({\cal F} \rho \| {\cal F} \sigma) \leq S(\rho \| \sigma)$.
\end{fact}

Jozsa~\cite{jozsa:fidelity} gave an elementary proof for finite
dimensional Hilbert spaces of the following basic and remarkable
property about fidelity.
\begin{fact}
\label{fact:jozsa}
Let $\rho, \sigma$ be density matrices in the same finite
dimensional Hilbert space $\hech$. Then for any finite dimensional
Hilbert space $\kay$ such that $\dim (\kay) \geq \dim (\hech)$, there
exist purifications
$\ket{\psi}, \ket{\phi}$ of $\rho, \sigma$ in $\hech \otimes \kay$,
such that
\begin{displaymath}
B(\rho, \sigma) = |\braket{\psi}{\phi}|.
\end{displaymath}
Also,
\begin{displaymath}
B(\rho, \sigma) = \trace{\sqrt{\rho} \sqrt{\sigma}}.
\end{displaymath}
\end{fact}

We will also need the following result about fidelity, proved
by Fuchs and Caves~\cite{fuchs:fidelity}.
\begin{fact}
\label{fact:fuchscaves}
Let $\rho, \sigma$ be density matrices in the same finite
dimensional Hilbert space $\hech$. Then
\begin{displaymath}
B(\rho, \sigma) = \inf_{F_1, \ldots, F_k} \sum_{i=1}^k
                        \sqrt{\Tr (F_i \rho) \, \Tr (F_i \sigma)},
\end{displaymath}
where $\{F_1, \ldots, F_k\}$ ranges over POVMs on $\hech$.
In fact, the infimum above can be attained by a complete orthogonal
measurement on $\hech$.
\end{fact}
The following relation is known between fidelity and trace distance between 
two density matrices ~\cite{nielsen:quant}.
\begin{fact}
\label{fact:Bandtrace}
Let $\rho, \sigma$ be density matrices in the same finite
dimensional Hilbert space $\hech$. Then
\begin{displaymath}
2(1 - B(\rho, \sigma)) \leq  \trace{\rho - \sigma } 
                    \leq 2\sqrt{1 - B(\rho, \sigma)^2}.
\end{displaymath}
\end{fact}

%We use $H(X)$ to denote the Shannon entropy of a classical
%random variable $X$.
The following information-theoretic facts follows easily from the
definitions.
\begin{fact}
\label{fact:inforelentropy}
Let $X$ be a classical random variable and
$M$ be a quantum encoding of $X$. Let $X$ take the values
$1, \ldots, l$ with probabilities $p_1, \ldots, p_l$ and
let $\sigma_1, \ldots, \sigma_l$ be
the respective density matrices of $M$. Let
$\sigma \defeq \sum_{j=1}^l p_j \sigma_j$ be the average density
matrix of $M$. Then, $I(X:M) = \sum_{j=1}^l p_j S(\sigma_j \| \sigma)$.
\end{fact}

\section{Improved Average encoding and Local transition theorem}     
\label{sec:impavglocal}

In this section, we observe that the following lemma
from~\cite{dacunha:BleqS} can be used to improve the average encoding
and local transition arguments of \cite{klauck:ptr}. If
Lemmas~\ref{lem:avgencoding} and \ref{lemappen:localtransition} are
used in their place, the factor $k^4$ in the denominator of some
existing lower bounds (e.g.~\cite{klauck:ptr} and \cite{jain:fsttcs})
can be replaced by $k^2$. 
\begin{lemma}
\label{lem:BleqS}
Let $\rho$ and $\sigma$ be two density matrices such that $S(\rho \| \sigma)$ 
is finite. Then, 
\begin{displaymath}
B(\rho , \sigma) \geq 2^{-S(\rho \| \sigma)/2}.
\end{displaymath}
\end{lemma}
\begin{proof}
Let $M$ be the complete orthogonal measurement which achieves the infimum 
as in the Fact~\ref{fact:fuchscaves}. Let $P$ and $Q$ be the 
classical distributions resulting after the measurement $M$ is performed. 
From Fact~\ref{fact:monorelentropy} and concavity of the $\log$ 
function it follows that:
\begin{eqnarray*}
-(1/2) S(\rho\|\sigma)\leq -(1/2)S(P \|Q) 
&=& \sum_i p_i \log  \sqrt{q_i/p_i} \\
& \leq & \log \sum_i \sqrt{q_i p_i} \\
&  = &   \log B(P, Q) = \log B(\rho,\sigma).
\end{eqnarray*}
\end{proof}
\begin{corollary}
\label{cor:BleqS}
Let $\rho$ and $\sigma$ be two density matrices such that $S(\rho \|
\sigma)$ is finite. Then,
\begin{displaymath}
1 - B(\rho , \sigma) \leq ((\ln 2)/ 2) S(\rho \| \sigma).
\end{displaymath}
\end{corollary}
\begin{proof}
If $((\ln 2)/ 2) S(\rho \| \sigma) \geq 1 $ then the inequality is 
trivial since $B(~,~) \geq 0$. Therefore when 
$((\ln 2)/ 2) S(\rho \| \sigma) \leq 1$, 
\begin{eqnarray*}
      B(\rho , \sigma) & \geq & 2^{-S(\rho \| \sigma)/2} \\   
                       & \geq & \exp ^{-((\ln 2)/2)S(\rho \| \sigma)}  \\
                       & \geq & 1 -((\ln 2)/2)S(\rho \| \sigma) 
~~(\mathrm{since} ~~ \exp ^{-x} \geq 1 -x, ~~ \mathrm{for} 
~~ 0 \leq x \leq 1) \\
\Rightarrow 1 - B(\rho , \sigma) & \leq & ((\ln 2)/2) S(\rho \| \sigma).
\end{eqnarray*}
\end{proof}

The following lemma follows immediately from the above corollary and 
Fact~\ref{fact:inforelentropy}.
\begin{lemma}[Average encoding theorem]
\label{lem:avgencoding}
Suppose $X$, $Q$ are two disjoint quantum systems, 
where $X$ is a classical
random variable which takes value $x$ with probability $p_x$, and
$Q$ is a quantum encoding $x \mapsto \sigma_x$ of $X$. Let the density
matrix of the average encoding be $\sigma \defeq \sum_x p_x \sigma_x$. 
Then,
\begin{displaymath}
\sum_x p_x (1- B(\rho, \rho_x)) \leq (\ln 2 / 2) I(X:Q).
\end{displaymath}
\end{lemma}

The following lemma follows immediately from Fact~\ref{fact:jozsa} 
and Fact~\ref{fact:Bandtrace} and Corollary~\ref{cor:BleqS}
\begin{lemma}[Local transition theorem]
\label{lemappen:localtransition}
Let $\rho_1, \rho_2$ be two density matrices in the same finite
dimensional Hilbert
space $\hech$, $\kay$ any Hilbert space of dimension at least
the dimension of $\hech$, and $\ket{\phi_i}$ any purifications
of $\rho_i$ in $\hech \otimes \kay$. Then, there is a local
unitary transformation $U$ on ${\cal K}$ that maps $\ket{\phi_2}$
to $\ket{\phi_2'} \defeq (I \otimes U) \ket{\phi_2}$ ($I$ is the
identity operator on $\hech$) such that
\begin{displaymath}
\trace{\ketbra{\phi_1} - \ketbra{\phi_2'}} 
        \leq 2\sqrt{1 - B(\rho_1, \rho_2)^2} 
        \leq 2\sqrt{2(1 - B(\rho_1, \rho_2))} 
         \leq 2\sqrt{\ln 2(S(\rho_1 \| \rho_2))}.
\end{displaymath}
\end{lemma}

\begin{fact}[\cite{lin}]
\label{fact:linclassical}
Suppose $X$ and $Q$ are two classical correlated random variables, where $X$ 
is uniformly distributed over $\{0,1\}$ and $Q$ is an
encoding $x \rightarrow P_x$ of $X$. Then, 
\[ 1 - B(P_1, P_2) \leq  I(X:Q). \]
\end{fact}

Following corollary is immediate from Fact~\ref{fact:fuchscaves} and 
monotonicity of information, 
\begin{corollary}
\label{cor:linquantum}
Suppose $X$ and $Q$ are disjoint quantum systems, where $X$ is a classical 
random variable uniformly distributed over $\{0,1\}$ and $Q$ is a quantum 
encoding $x \rightarrow \sigma_x$ of $X$. Then, 
\(1 - B(\sigma_1, \sigma_2) \leq  I(X:Q). \)
\end{corollary}

\end{document}